\documentclass[11pt]{amsart}
\usepackage{amsmath}
\usepackage{amssymb}
\usepackage{amscd}
\usepackage{epsfig}
\usepackage{amsxtra}
\usepackage{pifont}
\usepackage{mathabx}

\newcommand{\supp}{\mbox{\rm supp}}
\newcommand{\dens}{\mbox{\rm dens}}
\newcommand{\Vol}{\mbox{\rm Vol}}
\newcommand{\R}{{\mathbb R}}

\newcommand{\Z}{{\mathbb Z}}

\newcommand{\mig}{{\mathcal M}^\infty}

\newcommand{\hG}{\widehat{G}}

\newcommand{\SAP}{{\mathcal SAP}}
\newcommand{\WAP}{{\mathcal WAP}}

 \newtheorem{theorem}{Theorem}[section]
 \newtheorem{lemma}[theorem]{Lemma}
 \newtheorem{prop}[theorem]{Proposition}
 \newtheorem{cor}[theorem]{Corollary}

 \newtheorem{defi}[theorem]{Definition}
 
 \newtheorem{remark}[theorem]{Remark}

\begin{document}
\title[Weighted Model Combs]
{On Weighted Dirac Combs Supported Inside Model Sets}
\author{Nicolae Strungaru}
\address{Department of Mathematical Sciences, Grant MacEwan University
\\
10700 – 104 Avenue, Edmonton, AB, T5J 4S2;\\
and \\
Institute of Mathematics ``Simon Stoilow'' \\
Bucharest, Romania}
\email{strungarun@macewan.ca}
 \maketitle

\begin{abstract} In this paper we prove that given a weakly almost periodic measure $\mu$ supported inside some model set $\Lambda(W)$ with closed window $W$, then the strongly almost periodic component $\mu_S$ and the null weakly almost periodic component $\mu_0$ are both supported inside $\Lambda(W)$. As a consequence we prove that given any translation bounded measure $\omega$, supported inside some model set, then each of the pure point diffraction spectrum $\widehat{\gamma}_{pp}$ and the continuous diffraction spectrum $\widehat{\gamma}_c$ is either trivial or have a relatively dense support.
\end{abstract}

\section{Introduction}

The discovery of quasi-crystals in 1984 forced the International Union of Crystallography to amend the definition of crystal. By the new definition, "by "crystal" is meant any solid having an essentially discrete diffraction diagram, and by "aperiodic crystal" is meant any crystal in which three dimensional lattice periodicity can be considered to be absent."

Mathematically, physical diffraction is described as follows: given a translation bounded measure $\mu$ in a $\sigma$-compact locally compact Abelian group $G$, we construct a new measure $\gamma$ called the autocorrelation measure of $\mu$. As $\gamma$ is a positive definite measure, it is a Fourier Transformable measure, and its Fourier Transform $\widehat{\gamma}$ is a positive measure on $\hG$. The measure $\widehat{\gamma}$ describes the physical diffraction of $\mu$. As any measure, the diffraction has a Lesbegue decomposition with respect to the Haar measure $\theta_{\hG}$ of $\widehat{G}$:

$$\widehat{\gamma}=\widehat{\gamma}_{pp}+\widehat{\gamma}_{c}=\widehat{\gamma}_{pp}+\left( \widehat{\gamma}_{ac}+\widehat{\gamma}_{sc} \right) \,.$$

\noindent The measures $\widehat{\gamma}_{pp}, \widehat{\gamma}_{c}$ are called the discrete (or pure point) respectively continuous spectra of $\mu$.

It is usually understood that the diffraction is essentially discrete if the discrete spectrum $\widehat{\gamma}_{pp}$ has relatively dense support, that is if there exists some compact set $K \subset \hG$ so that $\widehat{\gamma}_{pp}(x+K) \neq 0$ for all $x \in \hG$.

The best mathematical models for quasicrystals are produced by cut and project schemes. This method was introduced by Y. Meyer in \cite{Meyer} in order to generate a class of models with interesting harmonic properties. The method was rediscovered independently by Kramer \cite{Kra} in 1984. Kramer used this method to produce a three dimensional icosahedral quasicrystal by projection from a 6 dimensional hypercubic lattice. It was only later that Lagarias \cite{LAG1} and Moody \cite{RVM3} connected the work of Meyer, Kramer and de Bruijn \cite{deB}, and realized the importance of this method to long range aperiodic order.
 
Given a locally compact group $H$, and a lattice $\widetilde{L} \subset G \times H$ such that the projection $p_G: \widetilde{L} \to G$ is one to one and the projection $p_H : \widetilde{L} \to H$ is dense,  then any compact set $W$ with non-empty interior defines a Delone set

$$\Lambda(W):= \{ x| (x,x^*) \in \widetilde{L}, x^* \in W \} \,.$$
The set $\Lambda(W)$ is called a model set. If $\Lambda(W)$ is regular model set, that is if the boundary of $W$ has Haar measure zero, then its diffraction measure $\widehat{\gamma}$ is purely discrete \cite{Martin2}.

While regular model sets are good models for quasi-crystals, they are usually hard to characterize. In contrast, in $\R^d$, there are few simple characterizations for relatively dense subsets of model sets \cite{Meyer}, \cite{LAG1}, \cite{RVM1}, and some of them have been extended recently to the case $G$ a $\sigma$-compact, locally compact Abelian group \cite{BLS}.

The simplest, due to Lagarias \cite{LAG1}, is the following: A relatively dense set $\Lambda \subset \R^d$ is a dense subset of a model set if and only if
$$\Delta:=\Lambda-\Lambda = \{ x-y |x,y \in \Lambda \} \,,$$
is uniformly discrete. Such a set is called a {\bf Meyer set}.

In \cite{NS1} we proved that for a Meyer set $\Lambda$, the discrete diffraction spectrum $\widehat{\gamma}_{pp}$ has a relatively dense support, and that the continuous spectrum $\widehat{\gamma}_{c}$ is either trivial or also has relatively dense support, a result which generalizes the similar results for subsets of lattices proven by M. Baake \cite{BA}, \cite{BGbook}. To prove this, we actually proved the stronger result that $\widehat{\gamma}_{pp}$ and $\widehat{\gamma}_{c}$ are strongly almost periodic measures, and that $\widehat{\gamma}_{pp}\neq 0$.

We also proved in \cite{NS2} that $\widehat{\gamma}_{pp}$ is a sup almost periodic measure, and that for some $C>0$ the characters in $\Delta^{C \epsilon}$ are also $\epsilon$ sup-almost periods of $\widehat{\gamma}$.

While the results in \cite{NS2} were easily extended to arbitrary translation bounded weighted Dirac combs supported inside model sets, the method of comparing the autocorrelation of the Meyer set to the autocorrelation of a larger regular model set we used in \cite{NS1} could not be extended beyond the case of real valued weighted combs \cite{NS1}, \cite{NS4}. The goal of this paper is to prove that in the case of weighted Dirac combs supported inside model sets, both the discrete and continuous diffraction spectra $\widehat{\gamma}_{pp}, \widehat{\gamma}_c$ are strongly almost periodic measures, a result which completes are previous work on the diffraction of measures supported on Meyer sets.

The reason why we could not extend the results of \cite{NS1}, \cite{NS4} to the case of complex valued weighted Dirac combs is simple: our key tool is the fact that the projection $P_S$ from the space of weakly almost periodic measures to the space of strongly almost periodic measures is positive thus reserves inequalities. Also, writing the complex weighted Dirac comb $\omega$ as a linear combination of two real values measures $\omega_1,\omega_2$ doesn't help either, as the autocorrelation of $\omega$ cannot be expressed in general as a linear combination of autocorrelations of $\omega_1$ and $\omega_2$.

What we do instead is express the autocorrelation $\gamma$ (or more generally any weakly almost periodic measure supported inside a fixed model set) as a linear combination of two real valued weakly almost periodic measures supported inside a model set and then try to use the properties of the projection $P_S$ on these two measures. In doing this, we cannot compare them anymore the autocorrelation of a regular model set, but we simply fix this by showing that we can always find a larger strongly almost periodic measure supported inside a model set.

The main result we get in this paper is the following:

\noindent {\bf Theorem \ref{main}} {\rm   Let $(G \times H, \widetilde{L})$ be any cut and project scheme, and let $W \subset H$ be compact. Then for all $\mu \in \WAP(G)$ with $\supp(\mu) \subset \Lambda(W)$, we have

$$\supp(\mu_S), \supp(\mu_0) \subset \Lambda(W) \,.$$}

This paper is organized as follows:

In Section \ref{Model S} we review the definitions and properties of cut and project schemes and model sets, while in Section \ref{WAP} we review the definitions and properties of almost periodic measures.

In Section \ref{domset} we define the new notion of dominating sets, and we study the connection between this notion and almost periodicity. We then use the results of this Section to prove Theorem \ref{main}.

We complete the paper by applying Theorem \ref{main} to the autocorrelation of an arbitrary weighted Dirac comb supported inside a model set. We combine this result with some previous results obtained in this direction to get:

\noindent{\bf Theorem \ref{diffapplic}} Let  $(G \times H, \widetilde{L})$ is a fixed cut and project scheme, $W \subset H$ is a pre-compact set and
$$\omega= \sum _{x \in \Lambda(W)} \omega(x) \delta_x \,,$$
be any translation bounded weighted Dirac comb supported inside $\Lambda(W)$. Let $\gamma$ be any autocorrelation of $\omega$.
Then,
\begin{itemize}
\item[i)] $\supp(\gamma_S), \supp(\gamma_0) \subset \Lambda(\overline{W-W})$.
\item[ii)] $\gamma_S$ is norm almost periodic.
\item[iii)]
$$\lim_{n} \frac{\left| \gamma_0 \right| (A_n)}{\Vol(A_n)} =0 \,.$$
\item[iv)] $(\widehat{\gamma})_{pp}, (\widehat{\gamma})_{c}$ are strongly almost periodic measure.
\item[v)] The set ${\mathcal B}:= \{ \chi | \widehat{\gamma}(\{\chi \}) \neq 0 \}$ of Bragg peaks is either empty or relatively dense.
\item[vi)] If $\omega \geq 0$ and there exists some $a>0$ so that $\{ x| \omega(x) >0\}$ is relatively dense, then ${\mathcal B}$ is relatively dense.
\item[vii)] The continuous spectra $\supp(\widehat{\gamma}_c)$ is either empty or relatively dense.
\item[viii)] There exists a $C>0$ so that for all $\epsilon >0$ and all $\chi \in \Lambda(\overline{W-W})^\epsilon$ and all $\psi \in \hG$ we have
$$\left|\widehat{\gamma}(\{ \psi + \chi\})-\widehat{\gamma}(\{ \psi \})  \right| \leq C \epsilon \,.$$
\item[ix)] $(\widehat{\gamma})_{pp}$ is a sup almost periodic measure.
\end{itemize}

\section{Model Sets}\label{Model S}

For the entire paper $G$ is a fixed $\sigma$-compact locally compact Abelian group.

In this section we review the notions of cut and project schemes, model sets, $\epsilon$-dual characters and dual cut and project scheme.

\begin{defi} A {\it cut and project scheme} consists of a direct
product $G \times H$ of $G$ and a locally compact Abelian group
$H$, and a lattice $\widetilde{L}$ in $G \times H$ such that with respect to
the natural projections $\pi_1 : G \times H \rightarrow G$ and
$\pi_2 : G \times H \rightarrow H$ we have:

\begin{itemize}
\item[i)] $\pi_1$ restricted to $L$ is one-to-one,
\item[ii)]$\pi_2(L)$
is dense in $H$.
\end{itemize}
\begin{equation} \label{S1cpScheme}
\begin{array}{ccccc}
 G & \stackrel{\pi_{1}}{\longleftarrow} & G\times H & \stackrel{\pi_{2}}
{\longrightarrow} & H   \quad .\\
 && \bigcup \\
 && \widetilde{L}
\end{array}
\end{equation}
We denote the cut and project scheme by $(G \times H , \widetilde{L})$.
\end{defi}

Let $(G \times H, \widetilde{L})$ be a cut and project scheme. Then, the restriction $\pi_1|_{\widetilde{L}} :  \widetilde{L} \to \pi_1( \widetilde{L})$ is a bijection, thus it has an inverse. Thus we can define a function $^* : \pi_1(\widetilde{L}) \to H$ by
$$x^*=\pi_2( \pi_1^{-1}(x) ) \,.$$
This function is called the {\bf star map}.

With this mapping, we get
$$\widetilde{L} = \{ (x,x^*)| x \in  \pi_1(\widetilde{L}) \} \,.$$
\begin{defi} A set $\Lambda \subset G$ is called a {\bf model set} if there exists a cut and project scheme $(G \times H, \widetilde{L})$ and a precompact set $W \subset H$  with non-empty interior, so that

 $$\Lambda= \Lambda(W) := \{ x \in \pi_1(\widetilde{L}) | x^* \in W \} \,.$$

\end{defi}

The following results were proven by Moody \cite{RVM1} in the case $G=\R^d$ and by Meyer \cite{Meyer} in the case of $H=\R^d$. We check now that their proofs can easily be extended beyond $\R^d$.

\begin{lemma}\label{LA1} {\rm Let $(G \times H, \widetilde{L})$ be a cut and project scheme, and let $W \subset H$.
\begin{itemize}
\item[i)] If $W$ has non-empty interior, then $\Lambda(W)$ is relatively dense.
\item[ii)]If $W$ is precompact, then $\Lambda(W)$ is uniformly discrete.
\end{itemize}}
\end{lemma}
\noindent{\bf Proof:}

\noindent{\bf (i):} The idea of the proof is simple: $\widetilde{L}+(K_1 \times K_2)= G \times H$. As $W$ has non-empty interior, $K_2$ can be covered by finitely many translates of $W$, and those translates can be chosen in $\pi_2(\widetilde{L})$. The preimage of those translates under the $*$-map defines a finite set $F$, and we will show that $\Lambda(W)$ is $-F+K_1$ relatively dense.

As $\widetilde{L}$ is relatively dense in $G \times H$, there exists $K_1, K_2$ be compact sets so that $\widetilde{L}+(K_1 \times K_2)= G \times H$. Also, as $\pi_2(\widetilde{L})$ is dense in $H$ and $W$ has non-empty interior, we have
$$K_2 \subset H= \pi_2(\widetilde{L})-W \,.$$
By the compactness of $K_2$, there exists a finite set $F \subset \pi_1(\widetilde{L})$ so that $K_2 \subset F^*-W$, where $F^* = \{ f^* | f \in F \}$. We claim that
$$\Lambda(W)+ \cup_{f \in F} (-f+K_1)= G \,.$$
Indeed, let $x \in G$. Then $(x,0) \in G \times H=\widetilde{L}+(K_1 \times K_2)$, thus we can write
$$(x,0)=(l,l^*)+(k,y) \, \mbox{with} \, (l,l^*) \in \widetilde{L} \,;\, (k,y) \in K_1 \times K_2 \,.$$
Since $y \in K_2 \subset F^*-W$, there exists some $f \in F$ and $w \in W$ so that $y=f^*-w$. Then
\begin{eqnarray*}
\begin{split}
(x,0)&=(l,l^*)+(k,y)=(x,0)=(l,l^*)+(k,0)+(0,f^*-w)\\
&=(l,l^*)+(k-f,0)+(f,f^*)+(0,-w)=(l+f,(l+f)^*)+(k-f,-w) \,, \\
\end{split}
\end{eqnarray*}
and hence
$$(x-k+f,w)=(l+f,(l+f)^*) \in \widetilde{L} \,.$$
This shows that $x-k+f \in \Lambda(W)$ and hence
$$x \in  \Lambda(W)+k-f \subset \Lambda(W)+K-F \,.$$
As $x \in G$ was arbitrary, we get that
$$G \subset  \Lambda(W)+(K-F) \,,$$
which proves that $\Lambda(W)$ is relatively dense.

\noindent{\bf (ii)}: Since $W$ is precompact, the set $\overline{W-W}$ is compact.
Let $0 \in U$ be a precompact open set in $G$. We will show that $\Lambda(\overline{W-W}) \cap U$ is finite, and thus we can find a smaller open set $0 \in V$ such that $\Lambda(\overline{W-W}) \cap V=\{ 0\}$. Thus $\Lambda(W)$ is $V$-uniformly discrete.

As $U \times \overline{W-W}$ is precompact in $G \times H$, and $\widetilde{L}$ is a lattice, the set
$$F:= \widetilde{L} \cap \left( U \times \overline{W-W} \right)$$
is finite. A straightforward computation yields
$$\Lambda(\overline{W-W}) \cap U =\pi_1(F) \,.$$
Thus $\Lambda(\overline{W-W}) \cap U$ is a finite set containing $0$. Then,  there exists $0 \in V \subset U$ open set in $G$ so that
$$\left[ \Lambda(\overline{W-W}) \cap U \right] \cap V = \{ 0 \} \,.$$
Since $V$ is a subset of $U$ we have
$$\{ 0 \}=\left[ \Lambda(\overline{W-W}) \cap U \right] \cap V = \Lambda(\overline{W-W}) \cap  V  \,.$$
Now our claim follows from
$$\Lambda(W)-\Lambda(W) \subset \Lambda(\overline{W-W}) \,.$$
\qed

We now introduce the notion of dual cut and project scheme. Our proof follows the proofs of \cite{RVM1}, \cite{Moody} very closely, we simply check that those proofs hold if we replace $\R^d$ by $G$.

\begin{lemma}\label{LA2} {\rm   Let $(G \times H, \widetilde{L})$ be a cut and project scheme, and let
$$\widetilde{L}^0:= \{ \chi \in \hat{G} \times \hat{H} | \chi(t)=1 \forall t \in  \widetilde{L} \,,$$
be the dual lattice. Then $(\hG \times \hat{H}, \widetilde{L}^0)$ is a cut and project scheme.
}
\end{lemma}
\noindent{\bf Proof:} First lets note that by \cite{Rudin} we have $\widehat{G \times H} \simeq \widehat{G} \times \widehat{H}$ and, under this isomorphism we have
$$\left( \frac{G \times H}{\widetilde{L}}\right)^{\wedge} \simeq \widetilde{L}^0 \,;\, \left( \frac{\hat{G} \times \hat{H}}{\widetilde{L}^0}\right)^{\wedge} \simeq \widetilde{L} \,.$$
Since $\frac{G \times H}{\widetilde{L}}$ is compact and $\widetilde{L}$ is discrete, by the Pontryagin Duality
$\widetilde{L}^0$ is discrete and $ \frac{\hat{G} \times \hat{H}}{\widetilde{L}^0}$ is compact, thus $\widetilde{L}^0$ is a lattice in $\hat{G} \times \hat{H}$.

We will denote the two projections of $\hat{G} \times \hat{H}$ onto $\hG$ and $\hat{H}$ by $\pi_1'$ and $\pi_2'$.

We start by showing that $\pi_2'(\widetilde{L}^0)$ is dense in $\hat{H}$:
\noindent By Pontryagin Duality, as the mapping $\pi_1 : \widetilde{L} \to G$ is one to one, the dual mapping
$$\widehat{\pi_1} : \hat{G} \to \frac{\hat{G} \times \hat{H}}{\widetilde{L}^0} $$
has dense image. We prove that this implies that the projection $\pi_2': \widetilde{L}^0 \to \hat{H}$ has dense image. Indeed, let $U \subset \hat{H}$ be any open set.
As $\widehat{\pi_1}$ has dense image, $\left( \hat{G} \times \{0\} + \widetilde{L}^0 \right) \cap \left( \hG \times U+\widetilde{L}^0 \right) \neq \emptyset$.
Let
$$(x,y)+ \widetilde{L}^0 \in \left( \hat{G} \times \{0\} + \widetilde{L}^0 \right) \cap \left(\hG \times U + \widetilde{L}^0 \right) \,.$$
Then, there exists two elements $l_1, l_2 \in \widetilde{L}^0$ such that
$$(x,y)+ l_1 \in \hat{G} \times \{0\} \,;\, (x,y)+ l_2 \in \hat{G} \times U \,.$$
Hence their difference $l_2 -l_1 \in \hat{G} \times U $ which proves that
$$\pi_2'(l_2-l_1) \in \pi_2'(\widetilde{L}^0) \cap U \,.$$
Hence $\pi_2'(\widetilde{L}^0) \cap U \neq \emptyset$. As $U$ was an arbitrary open set, this implies that $\pi_2'(\widetilde{L}^0)$ is dense in $H$.

Now we prove that the restriction $\pi_1'|_{\widetilde{L}^0}$ of $\pi_1'$ to $\widetilde{L}^0$ is one to one.

\noindent Again, by the Pontryagin Duality, as the mapping $\pi_2 : \widetilde{L} \to H$ had dense image, the dual mapping
$$\widehat{\pi_2} : \hat{H} \to \frac{\hat{G} \times \hat{H}}{\widetilde{L}^0} $$
is one to one. We show that this implies that $\pi_1'|_{\widetilde{L}^0}$ is one to one. Let $l \in \ker(\pi_1'|_{\widetilde{L^0}})$. As $\pi_1'(l) =0$, we can write $l=(0,y)$ with $y \in \hat{H}$. Then
$$\widehat{\pi_2}(y)= (0,y)+ {\widetilde{L}^0} = l+{\widetilde{L}^0}=0+{\widetilde{L}^0}=\widehat{\pi_2}(0) \,.$$
As $\widehat{\pi_2}$ is one to one, we get $l=0$, which shows that $\ker(\pi_1'|_{\widetilde{L^0}})=\{0 \}$.
\qed

We complete this section by introducing the $\epsilon$-dual characters and showing that for a model set $\Lambda(W)$ the set $\Lambda(W)^\epsilon$ is relatively dense.

\begin{lemma}\label{LA3} {\rm   Let $(G \times H, \widetilde{L})$ be a cut and project scheme, and let $W \subset H$ be precompact. Let $\epsilon >0$. Then the set
$$\Lambda(W)^\epsilon:= \{ \chi \in \hG | \left| \chi(x)-1 \right| \leq \epsilon \,$$ of {\bf $\epsilon$ dual characters} is relatively dense.
}
\end{lemma}
\noindent{\bf Proof}: Let $$N(\overline{W}, \epsilon)= \{ \chi \in \hat{H} | \left| \chi(x) -1 \right| < \epsilon \,;\, \forall x \in \overline{W} \} \,.$$
By Pontryagin duality,  as $\overline{W}$ is compact, the set $N(\overline{W}, \epsilon)$ is open in $\hat{H}$.

\noindent We will prove that $\Lambda(W)^\epsilon$ contains the set $\Gamma:=\Lambda(N(\overline{W}, \epsilon))$ defined in the dual cut and project scheme $(\hG \times \hat{H}, \widetilde{L}^0)$ by the window $N(\overline{W}, \epsilon)$. Then, by Lemma \ref{LA1}, (i), the set $\Lambda(W)^\epsilon$ is relatively dense.

\noindent Let $\Gamma$ be the model set defined in the dual cut and project scheme $(\hG \times \hat{H}, \widetilde{L}^0)$ by the window $N(\overline{W}, \epsilon)$, that is
$$\Gamma = \{ \psi \in \hG | \exists \chi \in N(\overline{W}, \epsilon) \mbox{so that}  (\psi, \chi) \in  \widetilde{L}^0 \} \,.$$
We show that $\Gamma \subset \Lambda(W)^\epsilon $. Indeed, let $\psi \in \Gamma$. Then, there exists $\chi \in N(\overline{W}, \epsilon)$ so that $(\psi,\chi) \in  \widetilde{L}^0 $.

\noindent Hence, for all $x \in \Lambda(W)$ we have $x^* \in W \subset \overline{W}$ and therefore
$$\left| \chi(x^*)-1 \right| < \epsilon \,.$$
Since $(x,x^*) \in  \widetilde{L} $ and $(\psi,\chi) \in  \widetilde{L}^0 $ we also have
$$\psi(x) \chi(x^*)=(\psi,\chi)(x,x^*)=1 \,.$$
Thus,
$$\left| \psi(x) -1 \right| = \left| \frac{1}{\chi(x^*)} -1 \right|= \left| \overline{\chi(x^*)} -1 \right|= \left| \overline{\chi(x^*) -1} \right| < \epsilon \,.$$
This shows that $\left| \psi(x) -1 \right| < \epsilon$ for all $x \in \Lambda(W)$,  and hence $\psi \in \Lambda(W)^\epsilon $.
\qed

\section{Almost Periodic Measures}\label{WAP}

In this section we review the definitions and properties of almost periodic measures, as introduced in \cite{ARMA}.

\begin{defi} A measure $\mu$ on $G$ is called {\bf translation bounded} (or {\bf shift bounded}) if for all $f \in C_C(G)$ the function
$f*\mu$ is uniformly continuous and bounded.

\noindent We will denote by $\mig(G)$ the space of translation bounded measures on $G$.
\end{defi}

All the measures we study in this paper are translation bounded.

\noindent It was proven in (\cite{ARMA1}, Theorem 1.1) that a measure $\mu$ is translation bounded if and only if, for each compact set $K$ there exists a constant $C_K$ such that
\begin{equation}\label{EQtb}
\left| \mu \right|(x+K) < C_K \,;\, \forall x \in G,
\end{equation}
where $ \left| \mu \right|$ denotes the variation od $\mu$. Baake and Moody \cite{BM} showed that this is equivalent to (\ref{EQtb}) holding for a single compact set with non-empty interior.

We now introduce the notion of almost periodicity. As almost periodicity for measures is obtained from almost periodicity for functions by convolution, we introduce first the notions of strongly and weakly almost periodicity for functions.

\begin{defi} A function $f \in C_U(G)$  is called Bohr-almost periodic if the set $\{ T^xf | x \in G \}$ is precompact in $(C_U(G), \| \, \|_\infty)$, where
$$T^xf(y) =f(-x+y ) \,,$$ and $C_U(G)$ denotes the space of uniformly continuous, bounded functions on $G$.
$f \in C_U(G)$  is called weakly almost periodic if the set $\{ T^xf | x \in G \}$ is precompact in the weak topology of $(C_U(G),\| \, \|_\infty )$.
\end{defi}

The space of weakly almost periodic functions is closed under complex conjugations and taking absolute values:

\begin{lemma}\label{overlinewap} {\rm (\cite{EBE}, Theorem 12.1)  If $f$ is weakly almost periodic, the $|f|$ and $\overline{f}$ are also weakly almost periodic functions.}
\end{lemma}

\noindent Eberlein proved that every weakly almost periodic measure is amenable \cite{EBE}, but since we don't need to use the notion of amenability in full, we simply introduce a simplified version of the mean $M(f)$ of a weakly almost periodic function $f$:

\begin{prop}{\rm (\cite{ARG}) Let $f$ be a weakly almost periodic function, and let $\{ A_n \}$ be a van Hove sequence (see for example \cite{BM}, Appendix for the definition). Then
$$\lim_n \frac{\int_{A_n} f(t) dt}{ \Vol(A_n)} \,,$$
exists. We will denote this limit by $M(f)$.}
\end{prop}
We are now ready to translate the concepts of almost periodicity from functions to measures.
\begin{defi} A translation bounded measure $\mu$  is called {\bf strongly almost periodic} if for all $f \in C_C(G)$, the function $\mu*f$ is a Bohr-almost periodic function.

\noindent $\mu$ is called {\bf weakly almost periodic} if for all $f \in C_C(G)$, the function $\mu*f$ is weakly almost periodic function.

\noindent $\mu$ is called {\bf null-weakly almost periodic} if for all $f \in C_C(G)$, the function $\mu*f$ is weakly almost periodic function and
$$M(\left| \mu*f \right|) = 0 \,.$$
We will denote by $\WAP(G), \SAP(G)$ and $\WAP_0(G)$ the subspaces (\cite{ARMA}, Proposition 5.1) of $\mig(G)$ consisting of weakly almost periodic measures, strongly almost periodic measures respectively null weakly almost periodic measures.
\end{defi}

The space $\WAP(G)$ contains all translation bounded positive definite measures. In Theorem \ref{decomp}, we will see that $\WAP(G)=\SAP(G) \bigoplus \WAP_0(G)$, and for double Fourier Transformable measures this decomposition is exactly the Fourier Dual decomposition of measures into discrete and continuous components (\cite{ARMA}, Theorem 11.2). Those two facts make the space $\WAP(G)$ and the decomposition below important in the study of mathematical diffraction.
\begin{theorem}\label{decomp}{\rm  (\cite{ARMA}, Theorem 7.2, Proposition 7.2 and Theorem 8.1 ) \rm  Every $\mu \in \WAP(G)$ can be written in an unique way as
$$\mu=\mu_S + \mu_0 \,;\, \mu_S \in \SAP(G), \mu_0 \in \WAP_0(G) \,.$$
Moreover, if $\mu \geq 0$ then $\mu_s \geq 0$.
}
\end{theorem}

We conclude the section by proving two simple results we will need in the remaining of the paper: Lemma \ref{L3} and Corollary \ref{MC1}.

\begin{lemma}\label{L3}{\rm   \rm Let $\mu \in \mig(G)$. Then $\mu$ is  weakly almost periodic measure if and only if $\overline{\mu}$ is  weakly almost periodic. }
\end{lemma}

\noindent{\bf Proof}: Since $\overline{\overline{\mu}}=\mu$ it suffices to prove that if $\mu \in \WAP(G)$ then $\overline{\mu} \in \WAP(G)$. We will see that follows immediately from the fact that this is true for functions:

\noindent Let $\mu \in \WAP(G)$ and let $f \in C_C(G)$ be arbitrary. Then $\overline{f} \in C_C(G)$ and thus $\mu*\overline{f} \in WAP(G)$. Then, by Lemma \ref{overlinewap} we get
$$\overline{\mu}*f=\overline{\mu*\overline{f}} \in WAP(G) \,.$$
This shows that $\overline{\mu}*f \in WAP(G)$ for all $f \in C_C(G)$.
\qed

\begin{remark}{\rm In \cite{ARMA}, the authors introduce a locally convex topology on $\mig$, which they call the product topology, and then define  weakly almost periodicity and strongly almost periodicity for a measure $\mu$ in terms of the set $\{  \delta_x * \mu| x \in G \}$ being precompact in the weak respectively product topology on $\mig$. Then, they prove in Corollary 5.5 that the definition is equivalent to the one we introduced before. It is easy to prove that the mapping $\mu \to \overline{\mu}$ is continuous in both the product and weak topology of $\mig$, which also can be used to prove Lemma \ref{L3}.}
\end{remark}

As we want to work with inequalities, we cannot work with complex valued measures. Instead we will look to the real and imaginary parts:
\begin{defi} \rm Let $\mu \in \mig(G)$. We define
$${\rm Re}(\mu)=\frac{1}{2}(\mu+\overline{\mu}) \,,$$
$${\rm Im}(\mu)=\frac{1}{2i}(\mu+\overline{\mu})\,.$$
\end{defi}
\noindent It is easy to check that if $\mu \in \mig$ then ${\rm Re}(\mu), {\rm Im}(\mu)$ are also translation bounded. We now show that they are real measures.

\begin{lemma}{\rm \label{L5}  \rm Let $\mu \in \mig(G)$. Then ${\rm Re}(\mu),{\rm Im}(\mu) $ are real valued measures.}
\end{lemma}
\noindent{\bf Proof}: Let $f \in C_C(G)$ be a real valued function. Then
$$\mu(f) + \overline{\mu(f)} \in \R \,.$$
Now, since $f$ is real valued, we have
$$\overline{\mu(f)}=\overline{\mu}(\overline{f})=\overline{\mu}(f) \,.$$
Thus
$$\left( \mu+ \overline{\mu} \right) (f) \in \R \,.$$
Similarly,
$$\frac{1}{2i} ( \mu - \overline{\mu})(f)=\frac{1}{2i} (\mu(f) - \overline{\mu(f)}) \in \R \,.$$
It follows that ${\rm Re}(\mu)(f), {\rm Im}(\mu)(f) \in \R$ for all real valued $f \in C_C(G)$, which proves our claim.
\qed

We next show that the real and imaginary parts of weakly almost periodic measures are also weakly almost periodic, which will allow us
to pass from complex valued weakly almost periodic measures to real valued weakly almost periodic measures when needed.

\begin{cor}{\rm \label{MC1}  \rm Let $\mu \in \mig(G)$. Then $\mu$ is  weakly almost periodic measure if and only if ${\rm Re}(\mu)$ and ${\rm Im}(\mu)$ are weakly almost periodic.}
\end{cor}
\noindent{\bf Proof}:

\noindent{\bf $\Rightarrow$}: This is an immediate consequence of Lemma \ref{L3}: $\overline{\mu}$ is also a weakly almost periodic measure and hence, as $\WAP(G)$ is a subspace of $\mig$ we get
$$\frac{1}{2}(\mu+\overline{\mu}) \,;\, \frac{1}{2i}(\mu-\overline{\mu}) \in \WAP(G) \,.$$

\noindent{\bf $\Leftarrow$}: Follows immediately from $\mu={\rm Re}(\mu)+i {\rm Im}(\mu)$.
\qed

We complete the section by introducing a large class of strongly almost periodic measures produced by a cut and project scheme.
\begin{lemma}{\rm (\cite{LR}, Theorem 3.1)\label{sap ms} Let $(G \times H, \widetilde{L})$ be a cut and project scheme and let $g \in C_C(H)$. Let
$$\omega_g := \sum{(x,x^*) \in \widetilde{L}} g(x^*) \delta_x \,.$$
Then, $\omega_g \in \SAP(G)$.  }
\end{lemma}

The result of Lemma \ref{sap ms} was first proven in the case $g=1_W*\widetilde{1_W}$ in (\cite{BM} Theorem 2 and Lemma 7), and it is easy to check that their proof works for arbitrary $g \in C_C(G)$. The result was later generalized to the case of rapidly decaying functions $g \in C_0(\R^d)$ \cite{CR}, and then to
rapidly decaying functions $g \in C_0(H)$ \cite{LR}. The class of measures of the type $\omega_g$ with $g \in C_C(H)$ is also studied in more detail and characterized in terms of norm almost periodicity in \cite{BLS}.

\section{Dominating sets}\label{domset}

Let $A \subset G$. We will denote by $\WAP(A), \SAP(A)$ and $\WAP_0(A)$ the spaces of weakly, strongly respectively null weakly almost periodic measures supported inside $A$. We define now a new notion of dominant sets:
\begin{defi}\label{domdef} Let $A$ be a closed set in $G$. We say that $B$ is a {\bf dominant set for $A$} if $A \subset B$ and for each real valued $\mu \in \WAP(A)$, there exists some $\nu \in \SAP(B)$ so that $\mu \leq \nu$.

\noindent We say that $B$ is a {\bf strong dominant set for $A$} if $A \subset B$ and for each real valued $\mu \in \mig(G)$ with $\supp(\mu) \subset A$, there exists some $\nu \in \SAP(B)$ so that $\mu \leq \nu$.
\end{defi}

\noindent \begin{remark} {\rm \begin{itemize} \item[i)] Let $A$ be a closed set in $G$ and let $B$  be a dominant set for $A$. If $C \subset A$ is closed and $B \subset D$ then $D$ is a dominant set for $C$.
\item[ii)] Let $A$ be a closed set in $G$. Then any strong dominant set for $A$ is also a dominant set for $A$.
\item[iii)] By replacing the measure $\mu$ in the definition of dominant set by $-\mu$, it can be shown that if $B$ is a dominant set for $A$ and $\mu \in \WAP(A)$, then there exists some measure $\nu' \in SAP(G)$ such that $\nu' \leq \mu$.

\noindent     The inequalities can also be inverted in the definition of strong dominant sets.
    \end{itemize}}
\end{remark}

\noindent In this paper we will only be interested in dominant sets for model sets. In the next Lemma, we show that given a model set $\Lambda(W)$, or more generally an  arbitrary uniformly discrete set $A$, there is a simple criteria which tells us if some set $B$ is a strong dominating set.

\begin{lemma}\label{L2}{\rm  Let $A$ be an uniformly discrete set in $G$ and let $A \subset B$. Then $B$ is a strong dominant set for $A$ if and only if there exists some $\nu \in \SAP(B)$ so that $\nu \geq \delta_A$.}
\end{lemma}
\noindent{\bf Proof}: \noindent{\bf $\Rightarrow$:} Follows immediately from the definition of strong dominant sets, as for uniformly discrete sets $A$ we have $\delta_A \in \mig(G)$.

\noindent{\bf $\Leftarrow$:} Let $\mu \in \mig(G)$ be a real value measure with $\supp(\mu)\subset A$. As $A$ is uniformly discrete, we can write
$$\mu=\sum_{x \in A} \mu(x) \delta_x \,.$$
Since $\mu$ is translation bounded, there exists some $C >0$ so that $\mu(x) < C \, \forall x \in A$. Then
$$\mu \leq C \delta_A \leq C \nu \,,$$
which completes the proof.
\qed

Combined with Lemma \ref{sap ms}, Lemma \ref{L2} shows that given a model set $\Lambda(W)$ we can always find a larger model set in the same cut and project scheme, which strongly dominates $\Lambda(W)$. This together with Theorem \ref{T1} below will prove our main result.
\begin{theorem}\label{T1} {\rm  Let $A$ be a closed set in $G$ and let $B$  be a dominant set for $A$. Then for all $\mu \in \WAP(A)$ we have
$$\supp(\mu_S), \supp(\mu_0) \subset B \,.$$}
\end{theorem}
\noindent{\bf Proof}: Since $\supp(\overline{\mu})=\supp(\mu)$, we get $\supp({\rm Re}(\mu)), \supp({\rm Im}(\mu))\subset A$.  Thus, by Lemma \ref{MC1}, it suffices to prove this result for real measures.

Let $\mu \in \WAP(A)$ be a real valued measure. Since $B$ dominates $A$, we can find two measures $\nu_1, \nu_2 \in {\mathcal SAP}(B)$ so that
$$\mu \leq \nu_1 \,;\, -\mu \leq \nu_2 \,.$$
Hence
$$\nu_1 - \mu \geq 0 \,;\, \nu_2 + \mu \geq 0 \,.$$
Now, we look at the projections on $\SAP(G)$. By Theorem \ref{decomp} we have
$$(\nu_1-\mu)_S \geq 0 \,;\, (\nu_2 + \mu)_S \geq 0 \,.$$
Using the linearity of the projection on the strongly almost periodic component, and the fact that by the uniqueness of the decomposition we have $\nu_i \in \SAP(G) \Rightarrow (\nu_i)_S =\nu_i$ we get
$$-\nu_2 \leq \mu_S \leq \nu_1  \,.$$
Now, as $\supp(\nu_1), \supp(\nu_2) \subset B$ we get
$$\supp(\mu_S) \subset B \,.$$
To complete the proof we just observe that since $\mu_0 = \mu-\mu_S$ and $\supp(\mu_S) \subset B \,,\, \supp(\mu) \subset A \subset B$ we get
$$\supp(\mu_0) \subset B \,.$$
\qed

Next, we show that if in a given cut and project scheme, model sets with pre-compact window are always dominated by model sets with larger open window. This is an immediate consequence of Lemma \ref{sap ms}, Lemma \ref{L2} and the fact that if $W \subset U$ are so that $U$ is open and $W$ compact, then we can find a non-negative function $f \in C_C(G)$ which is 1 on $W$ and 0 outside $U$.

\begin{prop}\label{T2} {\rm   Let $(G \times H, \widetilde{L})$ be any cut and project scheme, and let $W \subset H$ be pre-compact. Let $U$ be any open set containing $W$.
Then $\Lambda(U)$ is a strong dominating set for $\Lambda(W)$.}
\end{prop}
\noindent{\bf Proof}: Let $g \in C_C(H)$ be so that $g \geq 0, g(x)=1 \forall x \in H$ and $\supp(g) \subset U$.  Then, by Lemma \ref{sap ms} we have $\omega_g \in \SAP(G)$. In particular, $\omega_g \in \SAP(\Lambda(U))$. As $g \equiv 1$ on $W$ we get
$$\omega_g \geq \delta_{\Lambda(W)} \,,$$  and the result follows from Lemma \ref{L2}.
\qed

We now proceed to prove the main result in the paper.

\begin{theorem}\label{main}{\rm   Let $(G \times H, \widetilde{L})$ be any cut and project scheme, and let $W \subset H$ be compact. Then for all $\mu \in \WAP(\Lambda(W))$ we have
$$\supp(\mu_S), \supp(\mu_0) \subset \Lambda(W) \,.$$}
\end{theorem}
\noindent{\bf Proof:} Let $U \subset H$ be a fixed precompact open set containing $W$. Then by Proposition \ref{T2} the set $\Lambda(U)$ is a strong dominating set for $\Lambda(W)$, and thus, by Theorem \ref{T1},  $\supp(\mu_S) \subset \Lambda(U)$. We complete the proof by showing that $\mu_S$ is zero at every point in $\Lambda(U)\backslash \Lambda(W)$.

We know that $\Lambda(U)$ is uniformly discrete. Thus, since $\supp(\mu_S) \subset \Lambda(U)$ we can write
$$\mu_S=\sum_{x \in \Lambda(U)} \mu_S(\{ x \}) \delta_x \,.$$
Let $x \in \Lambda(U) \backslash \Lambda(W)$. We claim $\mu_S(\{ x \})=0$.

\noindent As $x \in \Lambda(U) \backslash \Lambda(W)$ we have $x^* \in (U_0 \backslash W)$. We define $U_x:= U \backslash \{ x^* \}$. Then $U_x$ is open in $H$ and $W \subset U_x$. Then, by combining again Proposition \ref{T2} and Theorem \ref{T1}, we get $\supp(\mu_S) \subset \Lambda(U_x)$ and hence $\mu_S(\{ x \})=0$ as desired.
\qed

\begin{remark} {\rm Under the assumptions of Theorem \ref{main}, if $W$ has empty interior, by exactly the same argument as in the proof of (\cite{BM}, Lemma 4) one can show that $\Lambda(W)$ is not relatively dense. In this case it follows that $\SAP(\Lambda(W))=\{ 0\}$. In particular, Theorem \ref{main} implies that any measure $\mu \in \WAP(\Lambda(W))$
is automatically null weakly almost periodic.}
\end{remark}

Combining Theorem \ref{main} with Theorem \ref{decomp} we get:

\begin{cor}\label{CQ1}{\rm   Let $(G \times H, \widetilde{L})$ be any cut and project scheme, and let $W \subset H$ be compact, with non-empty interior. Let $\Lambda=\Lambda(W)$. Then
$$\WAP(\Lambda)=\SAP(\Lambda)+\WAP_0(\Lambda) \,.$$}
\end{cor}

\section{On the autocorrelation and diffraction for weighted model combs}

In this section we look at the consequences of Theorem \ref{main} to the diffraction of weighted Dirac combs supported inside model sets.
We skip the formal definitions of autocorrelation and diffraction measures, and refer the reader instead to \cite{BG}.

Given a model set $\Lambda(W)$ in some cut and project scheme $(G \times H, \widetilde{L})$ and some translation bounded measure
$$\omega=\sum_{x \in \Lambda(W)} \omega(x) \delta_x \,,$$
then any autocorrelation $\gamma$ of $\omega$ can be written \cite{BM}
$$\gamma=\sum_{x \in \Lambda(W-W)} \gamma(x) \delta_x \,.$$
As $\gamma$ is a weakly almost periodic measure supported inside $\Lambda(\overline{W-W})$, we can apply Theorem \ref{main} to this measure to obtain $$\supp(\gamma_S), \supp(\gamma_0) \subset \Lambda(\overline{W-W}) \,,$$
which leads to some interesting properties of the discrete and continuous spectra of $\omega$.

We combine this together with some results proven in \cite{BM}, \cite{NS1}, \cite{NS2} and \cite{NS4} in the Theorem \ref{diffapplic}:

\begin{theorem}\label{diffapplic} Let $(G \times H, \widetilde{L})$ be a cut and project scheme, $W \subset H$ be a pre-compact set and
$$\omega= \sum _{x \in \Lambda(W)} \omega(x) \delta_x \,,$$
be any translation bounded weighted Dirac comb supported inside $\Lambda(W)$. Let $\gamma$ be any autocorrelation of $\omega$, and $A_n$ be any van Hove sequence in $G$. Then
\begin{itemize}
\item[i)] $\supp(\gamma_S), \supp(\gamma_0) \subset \Lambda(\overline{W-W})$.
\item[ii)] $\gamma_S$ is norm almost periodic.
\item[iii)]
$$\lim_{n} \frac{\left| \gamma_0 \right| (A_n)}{\Vol(A_n)} =0 \,.$$
\item[iv)] $(\widehat{\gamma})_{pp}, (\widehat{\gamma})_{c}$ are strongly almost periodic measure.
\item[v)] The set ${\mathcal B}:= \{ \chi | \widehat{\gamma}(\{\chi \}) \neq 0 \}$ of Bragg peaks is either empty or relatively dense.
\item[vi)] If $\omega \geq 0$ and there exists some $a>0$ so that $\{ x| \omega(x) >0\}$ is relatively dense, then ${\mathcal B}$ is relatively dense.
\item[vii)] The continuous spectra $\supp(\widehat{\gamma}_c)$ is either empty or relatively dense.
\item[viii)] There exists a $C>0$ so that for all $\epsilon >0$ and all $\chi \in \Lambda(\overline{W-W})^\epsilon$ and all $\psi \in \hG$ we have
$$\left|\widehat{\gamma}(\{ \psi + \chi\})-\widehat{\gamma}(\{ \psi \})  \right| \leq C \epsilon \,.$$
\item[ix)] $(\widehat{\gamma})_{pp}$ is a sup almost periodic measure.
\end{itemize}
\end{theorem}

\noindent{\bf Proof:} {\bf i):} As $W$ is precompact, the set $\overline{W-W}$ is compact, and thus $\Lambda-\Lambda$ is uniformly discrete. Then , as
$$\supp(\gamma) \subset \Lambda - \Lambda \subset \Lambda(\overline{W-W}) \,,$$
the claim follows from Theorem \ref{main}.

\noindent {\bf ii)}:  follows from (\cite{BM}, Proposition 8).

\noindent {\bf iii):} is proven in (\cite{NS1}, Proposition 5.7).

\noindent {\bf iv):} is a consequence of (\cite{ARMA}, Corollary 11.1).

\noindent {\bf v)} and {\bf vii):} follow from (\cite{NS1}, Proposition 3.5).

\noindent {\bf vi):}  is a consequence of ( \cite{NS4}, Theorem 4.9).

\noindent We now complete the proof by proving {\bf viii)} and {\bf ix)}. The proof is almost identical to (\cite{NS2}, Proposition 9.4), the only difference is that by a simple argument we eliminate the requirement that $\gamma$ is positive.

\noindent As $\omega$ is translation bounded, we get that $\gamma$ is translation bounded. Thus, there exists a constant $D>0$ do that
$$\left| \gamma(\{ x \}) \right| \leq D \,;\, \forall x \in D \,.$$
Let $\Lambda(K)$ be any regular model set containing $\Lambda(\overline{W-W})$. Then, as $\left| \gamma \right| \leq D \delta_{\Lambda(K)}$ we have
$$\frac{\left| \gamma \right|(A_n)}{\Vol(A_n)} \leq D \frac{ \sharp(\Lambda(K) \cap A_n)}{\Vol(A_n)} \,.$$
By (\cite{RVM1}, Theorem 1)
$$\dens(\Lambda(K))=\lim_n  \frac{ \sharp(\Lambda(K) \cap A_n)}{\Vol(A_n)} \,.$$ exists and is finite, thus
$$C:= \limsup_{n \to \infty} \frac{\left| \gamma \right|(A_n)}{\Vol(A_n)} < \infty \,.$$
Now, by \cite{DL02}, for all $\varphi \in \hG$ we have
$$\widehat{\gamma}(\{ \varphi \})= \lim_n \frac{ (\overline{\varphi} \gamma)(A_n)}{\Vol(A_n)} \,.$$
Thus

\begin{eqnarray}
\begin{split}
&\left|\widehat{\gamma}(\{ \psi + \chi\})-\widehat{\gamma}(\{ \psi \})  \right|= \left| \lim_n \frac{ (\overline{ \psi + \chi} \gamma-\overline{ \psi} \gamma)(A_n)}{\Vol(A_n)} \right| \\
&\leq \limsup_n \frac{ \left| (\overline{ \psi + \chi} \gamma-\overline{ \psi } \gamma)(A_n) \right| }{\Vol(A_n)} \\
&\leq \limsup_n \frac{ \left| (\overline{ \psi + \chi}-\overline{ \psi })\right| \left|\gamma\right| (A_n)  }{\Vol(A_n)} \\
&= \limsup_n \frac{\sum_{t \in \Lambda(\overline{W-W})\cap A_n} \left| \overline{ \psi + \chi}(t)-\overline{ \psi }(t)\right| \left|\gamma\right|(t) }{\Vol(A_n)} \\
&= \limsup_n \frac{\sum_{t \in \Lambda(\overline{W-W})\cap A_n} \left| \overline{ \psi}(t)\right|-\left|\overline{ \chi }(t)-1\right| \left|\gamma\right|(t)}{\Vol(A_n)} \\
&\leq \limsup_n \frac{\sum_{t \in \Lambda(\overline{W-W})\cap A_n} \epsilon \left|\gamma\right|(t)  }{\Vol(A_n)} \\
&= \epsilon \limsup_n \frac{\left|\gamma\right|(A_n) )  }{\Vol(A_n)} =\epsilon C  \,. \\
\end{split}
\end{eqnarray}
This proves $vii)$.  $ix)$ follows now from $viii)$ and the relatively denseness of $\Lambda(\overline{W-W})^\epsilon$ in $\hG$.
\qed

\begin{remark}  {\rm
If $\omega(x) \geq 0 \, \forall x \in \Lambda$, then $\gamma_S$ belongs to the cone ${\mathcal PSD}(G)$ of positive, positive definite, strongly almost periodic discrete measures introduced in \cite{NS4}.}
\end{remark}

We complete the section by proving  that the conditions $ii)$ and $iii)$ in Theorem \ref{diffapplic} determine the decomposition $\gamma=\gamma_S+\gamma_0$.

\begin{lemma}\label{l25}  {\rm  Let $\mu$ be a weakly almost periodic measure, let $A_n$ be a van Hove sequence and let $\mu_1, \mu_2$ be such that
\begin{itemize}
\item[i)]$\mu=\mu_1+\mu_2$,
\item[ii)] $\mu_1$ is norm almost periodic,
\item[iii)]
$$\lim_n \frac{\left| \mu_2 \right| (A_n)}{\Vol(A_n)} =0 \,.$$
\end{itemize}
Then
$$\mu_1= \mu_S \,;\, \mu_2 =\mu_0 \,.$$
}
\end{lemma}
\noindent{\bf Proof:} Since $\mu_1$ is norm almost periodic, it is strongly almost periodic (\cite{BM}, Lemma 7).
Therefore $\mu_2=\mu-\mu_1$ is weakly almost periodic as the difference of two weakly almost periodic measures. It follows from (\cite{NS1}, Proposition 5.7), $\mu_2$ is null weakly almost periodic.

\noindent Then the claim follows from  the uniqueness of Theorem \ref{decomp}.

\qed

\begin{remark}  {\rm We should emphasize here that while for measures supported inside model sets, the decomposition $\mu=\mu_s+\mu_0$ is uniquely characterized by Lemma \ref{l25}, for general weakly almost periodic measures, $\mu_s$ is not necessarily norm almost periodic and $\mu_0$ does not necessarily satisfy
$$\lim_n \frac{\left| \mu_0 \right| (A_n)}{\Vol(A_n)} =0 \,.$$
For example
$$\mu=\delta_{\Z}+\delta_{\left( \sqrt{2} \Z\right)+\frac{1}{2}}\,.$$
is a strongly almost periodic measure, but $\mu_s=\mu$ is not a norm almost periodic (\cite{NS2}, Proposition 7.3).

\noindent Also, as shown in (\cite{NS1}, Example 5.8)  the measure
$$\nu=\delta_{\Z}-\sum_{n \in \Z^*} \delta_{n-\frac{1}{n}} \,,$$ is a null weakly almost periodic measure, but
$$\lim_n \frac{\left| \nu \right| (A_n)}{\Vol(A_n)} =2 \,.$$
}
\end{remark}

\subsection*{Acknowledgments} I wish to thank  Dr. Michael Baake for reading the manuscript and for his suggestions which improved the quality of the paper.

\end{document}